\documentclass[aps,pra,preprint]{revtex4-1}
\usepackage{amsfonts,bm,graphicx,hyperref,physics,pifont,color}
\usepackage[caption=false]{subfig}

\definecolor{myblue}{rgb}{0.7,0.75,1}
\definecolor{orange}{rgb}{1,0.5,0}
\definecolor{mygreen}{rgb}{0,0.5,0}

\begin{document}
\title{\bf Variational Quantum Factoring}

\author{Eric R.\ Anschuetz}
\email{eric.anschuetz@zapatacomputing.com}
\affiliation{Zapata Computing Inc., 501 Massachusetts Avenue, Cambridge MA 02138}
\author{Jonathan P.\ Olson}
\email{jonny@zapatacomputing.com}
\affiliation{Zapata Computing Inc., 501 Massachusetts Avenue, Cambridge MA 02138}
\author{Al\'{a}n Aspuru-Guzik}
\email{aspuru@zapatacomputing.com}
\affiliation{Zapata Computing Inc., 501 Massachusetts Avenue, Cambridge MA 02138}
\author{Yudong Cao}
\email{yudong@zapatacomputing.com}
\affiliation{Zapata Computing Inc., 501 Massachusetts Avenue, Cambridge MA 02138}

\begin{abstract}

Integer factorization has been one of the cornerstone applications of the field of quantum computing since the discovery of an efficient algorithm for factoring by Peter Shor. Unfortunately, factoring via Shor's algorithm is well beyond the capabilities of today's noisy intermediate-scale quantum (NISQ) devices.
In this work, we revisit the problem of factoring, developing an alternative 
to Shor's algorithm, which employs established techniques to map the factoring problem to the ground state of an Ising Hamiltonian. The proposed variational quantum factoring (VQF) algorithm starts by simplifying equations over Boolean variables in a preprocessing step to reduce the number of qubits needed for the Hamiltonian. Then, it seeks an approximate ground state of the resulting Ising Hamiltonian by training variational circuits using the quantum approximate optimization algorithm (QAOA). 
We benchmark the VQF algorithm on various instances of factoring and present numerical results on its performance. 
\end{abstract}

\maketitle
\section{Introduction}

Integer factorization is one of the first practically relevant problems that can be solved exponentially faster on a quantum computer than any currently known methods for classical computation by employing Shor's factoring algorithm~\cite{Shor1999}. Since its initial appearance, numerous follow-up studies have been carried out to optimize the implementation of Shor's algorithm from both algorithmic and experimental perspectives~\cite{Lu2007DemonstrationQubits,Martin-Lopez2012ExperimentalRecycling,Lanyon2007ExperimentalEntanglement,Politi2010AChip,Lucero2012ComputingProcessor,Geller2013FactoringQubits,Beauregard2003CircuitQubits,Ekera2016ModifyingLogarithms,Ekera2017QuantumIntegers,Monz2016RealizationAlgorithm}. Improved constructions~\cite{Beauregard2003CircuitQubits,Haner2016FactoringMultiplication,Takahashi:2006:QCS:2011665.2011669} have been proposed which, for an input number of $n$ bits, improve the circuit size from $3n$ qubits~\cite{Nielsen2000QuantumInformation} to $2n+3$~\cite{Beauregard2003CircuitQubits} and $2n+2$~\cite{Haner2016FactoringMultiplication} qubits, and with nearest-neighbor interaction constraints~\cite{Fowler2004ImplementationArray}. It has also been pointed out that using iterative phase estimation~\cite{Kitaev1995QuantumProblem}, one can further reduce the qubit cost to $n+1$, though the circuit needs to be adaptive in this case~\cite{Monz2016RealizationAlgorithm,Martin-Lopez2012ExperimentalRecycling}. Various other implementations~\cite{Zalka2006ShorsQubits,Gidney2017FactoringQubits} of Shor's algorithm have been proposed such that only a subset of qubits need to be initialized in a computational basis state (``clean qubits'').

Concrete resource estimates in realizing Shor's algorithm for factoring relevant numbers for RSA have also been performed for specific architectures~\cite{Fowler2018ScalabilityGates,Jones2012LayeredComputing,RodneyVanMeterThaddeusD.LaddAustinG.Fowler2010DistributedNanophotonics,Thaker2006QuantumComputing}. For example, on one particular architecture of a fault-tolerant quantum computer~\cite{RodneyVanMeterThaddeusD.LaddAustinG.Fowler2010DistributedNanophotonics,Jones2012LayeredComputing} it is estimated that factoring a 2048-bit RSA number requires a circuit depth on the order of $10^9$, {requiring} roughly 10 days on a quantum computer {comprised} of $10^5$ logical qubits~\cite[{cf.\ }Figure 15]{Jones2012LayeredComputing}. Another resource estimate~\cite{Devitt2013RequirementsComputer} {considering a} photonic architecture suggests that factoring a 1024-bit RSA number would take 2.3 years with 1.9 billion photonic modules. In contrast, present technologies are in the era of noisy intermediate-scale quantum (NISQ) devices~\cite{Preskill2018QuantumBeyond}, where quantum devices typically have on the order of $10^2$-$10^3$ noisy qubits that can only reliably implement circuits of limited depth. This renders the practical impact of Shor's algorithm (as well as alternative algorithms for quantum factoring that use subroutines requiring fault tolerance, such as~\cite{BernsteinAAlgorithm,Ekera2017QuantumIntegers}) a reality at least as distant as the realization of fault-tolerant quantum computers. 

Another approach to factoring on a quantum computer exploits the mapping from factoring to the ground state problem of an Ising Hamiltonian~\cite{factoring-as-optimization}. The basic idea underlying the mapping is to simply use the fact that factoring is the inverse operation of multiplication. Therefore, by working through the multiplication of two undetermined $n$-bit numbers and fixing the output to be the number being factored, one can write a set of equations involving the bits of the factors and the carry bits. The Hamiltonian is constructed such that the ground state satisfies all of the generated equations and any bit assignment which violates any of the equations receives an energy penalty. Interesting observations~\cite{Dattani2014,Geller2013FactoringQubits,Xu2012QuantumSystem} have been made about specific instances of factoring which allow one to simplify the equations tremendously. On the experimental side, most of the current efforts focus on analog approaches such as quantum annealing~\cite{Schaller:2010:RSA:2011438.2011447,Jiang2018QuantumFactorization} and simulated adiabatic evolution~\cite{Peng2008,Xu2012QuantumSystem}. However, the same ground state problem of Ising Hamiltonians can be {approximately} solved on gate model NISQ devices using the quantum approximate optimization algorithm (QAOA)~\cite{Farhi2014}.

Here we introduce an approach which we call \emph{variational quantum factoring} (VQF). As with other hybrid classical/quantum algorithms such as the variational quantum eigensolver (VQE)~\cite{Peruzzo2014AProcessor.} or the quantum autoencoder (QAE)~\cite{Romero2016QuantumData}, classical preprocessing coupled with quantum state preparation and measurement are used to optimize a cost function. In particular, we employ the QAOA algorithm~\cite{Farhi2014} and classical preprocessing for factoring.
The VQF scheme has two main components: first, we map the factoring problem to an Ising Hamiltonian, using an automated program to find reduction in the number of required qubits whenever appropriate. Then, we train the QAOA ansatz for the Hamiltonian using a combination of local and global optimization. 
We explore six instances of the factoring problem (namely, the factorings of 35, 77, 1207, 33667, 56153, and 291311) to demonstrate the effectiveness of our scheme in certain regimes as well as its robustness with respect to noise. 

The remainder of the paper is organized as follows: Section~\ref{sec:encode} describes the mapping from {a} factoring {problem} to an Ising Hamiltonian, together with the simplification scheme that is used for reducing the number of qubits needed.
Section~\ref{sec:hybrid} introduces QAOA and describes our method for training the ansatz. Section~\ref{sec:sim} presents our numerical results. We conclude in Section~\ref{sec:discuss} with further discussion on future works.
\section{Encoding factoring into an Ising Hamiltonian}
\label{sec:encode}

\subsection{Factoring as binary optimization} It is known from previous work that factoring can be cast as the minimization of a cost function~\cite{factoring-as-optimization}, which can then be encoded into the ground state of an Ising Hamiltonian~\cite{Dridi2017,Dattani2014,Xu2012}. To see this, consider the factoring of $m=p\cdot q$, each having binary representations 
\begin{equation}
\begin{array}{ccl}
m&=&\displaystyle\sum_{k=0}^{n_m-1} 2^i m_k, \\[0.1in]
p&=&\displaystyle\sum_{k=0}^{n_p-1} 2^i p_k, \\[0.1in]
q&=&\displaystyle\sum_{k=0}^{n_q-1} 2^i q_k, 
\end{array}
\end{equation}
where $m_k\in\{0,1\}$ is the $k$th bit of $m$, $n_m$ is the number of bits of $m$, and similarly for $p$ and $q$. When $n_p$ and $n_q$ are unknown (as they are unknown \textit{a priori} when only given a number $m$ to factor), one may assume without loss of generality~\cite{factoring-as-optimization} that $p\geq q$, $n_p=n_m$, and $n_q=\left\lceil\frac{n_m}{2}\right\rceil$~\footnote{To lower the needed qubits for our numerical simulations, we assumed prior knowledge of $n_p$ and $n_q$.}.
By carrying out binary multiplication, the bits representing $m$, $p$, and $q$ must satisfy the following set of $n_c=n_p+n_q-1\in O(n_m)$ equations \cite{factoring-as-optimization,Xu2012,Dridi2017}:
\begin{equation}
0=\sum\limits_{j=0}^{i} q_j p_{i-j}+\sum\limits_{j=0}^i z_{j,i}-m_i-\sum\limits_{j=1}^{n_c} 2^j z_{i,i+j}
\label{factoring_zero}
\end{equation}
for all $0\leq i< n_c$, where $z_{i,j}\in\left\{0,1\right\}$ represents the carry bit from bit position $i$ into bit position $j$. If we associate a clause $C_i$ over $\mathbb{Z}$ with each equation such that
\begin{equation}
C_i=\sum\limits_{j=0}^{i} q_j p_{i-j}+\sum\limits_{j=0}^i z_{j,i}-m_i-\sum\limits_{j=1}^{n_c} 2^j z_{i,i+j},
\label{factoring_clause}
\end{equation}
then factoring can be represented as finding the assignment of binary variables $\left\{p_i\right\}$, $\left\{q_i\right\}$, and $\left\{z_{ij}\right\}$ which solves
\begin{equation}
0=\sum\limits_{i=0}^{n_c}C_i^2.
\label{classical_zero}
\end{equation}

In general, if $m$ contains more than two prime factors, Equation~\ref{factoring_zero} still holds and our method will produce a Hamiltonian with a ground state manifold degenerate over all pairs of factors of $m$. To further factor $m$, one can repeat the VQF scheme to possibly yield a different $\left(p,q\right)$ pair, recursively apply our scheme to each of $p$ and $q$, or expand $m$ into a product of multiple factors to arise at an analogous form of Equation~\eqref{factoring_zero} that can be simultaneously solved for all factors of $m$. If $m$ is prime itself, then its primality can be easily detected~\cite{Agrawal2004}. Therefore, for the rest of our discussion we will consider $m$ to be the product of two primes (a \emph{biprime}), without loss of generality.

\subsection{Simplifying the clauses}\label{sec:simplification}
One method for simplifying clauses is to directly solve for a subset of the binary variables that are easy to solve for classically~\cite{Xu2012,Dattani2014}. This reduction iterates through all clauses $C_i$ as given by Equation~\eqref{factoring_clause} a constant number of times. In the following discussion, let $x,y,x_i\in\mathbb{F}_2$ be unknown binary variables and $a,b\in\mathbb{Z}^+$ positive constants. 
Along with some trivial relations, we apply the classical preprocessing rules~\footnote{We note that other simple relations exist that can be used for preprocessing---the simplified clauses for $m=56153,291311$ as used in our numerical simulations were given by~\cite{Dattani2014} who utilized a different preprocessing scheme.}:
\begin{equation}
\begin{aligned}
xy-1=0&\implies x=y=1,\\
x+y-1=0&\implies xy=0,\\
a-bx=0&\implies x=1,\\
\sum_i x_i=0&\implies x_i=0,\\
\sum_{i=1}^a x_i-a=0&\implies x_i=1.
\end{aligned}
\label{boolean_subs}
\end{equation}

We also are able to truncate the summation of the final term in Equation~\eqref{factoring_clause}. This is done by noting that if $2^j$ is larger than the maximum attainable value of the sum of the other terms, $z_{i,i+j}$ cannot be one; otherwise, the subtrahend would be larger than the minuend for all possible assignments of the other variables, and Equation~\eqref{factoring_zero} would never be satisfied. This effectively limits the magnitude of Equation~\eqref{factoring_clause} to be $O\left(n_m\right)$.

\begin{figure}[ht]
\resizebox{\linewidth}{!}{
\includegraphics{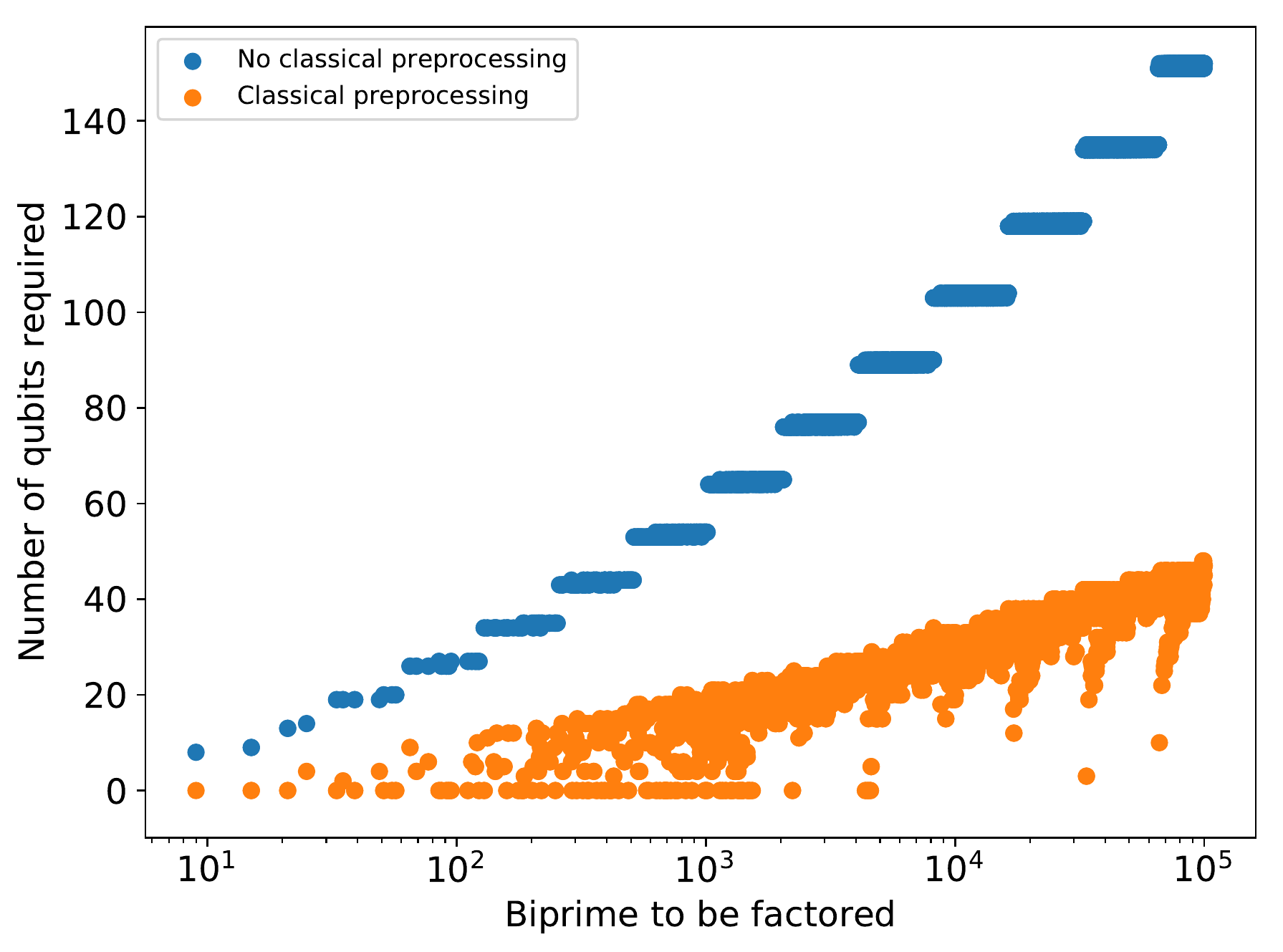}
}
\caption{This figure empirically demonstrates the reduction in qubit requirements after implementing the classical preprocessing procedure outlined in Section~\ref{sec:simplification}. After the classical preprocessing algorithm (orange), the number of qubits necessary for our algorithm empirically scales approximately as $O\left(n_m\right)$. In contrast, with no simplification (blue), VQF's qubit requirements scale as $O\left(n_m\log\left(n_m\right)\right)$ asymptotically~\cite{factoring-as-optimization}.}
\label{fig:preprocess}
\end{figure}

This classical preprocessing iterates through each of $O\left(n_c\right)$ terms in each of $n_c\in O\left(n_m\right)$ clauses $C_j$ (see Equations \ref{factoring_zero} and \ref{factoring_clause}), yielding a classical computer runtime of $O\left(n_m^2\right)$. This is because $O\left(n_c\right)=O\left(n_m\right)$ from the identity $n_c=n_p+n_q-1$, and $n_p\leq n_m$ and $n_q\leq\lceil\frac{n_m}{2}\rceil$~\cite{factoring-as-optimization}. In practice, for most instances we have observed that the preprocessing program greatly reduces the number of (qu)bits needed for solving the problem, as is shown in Figure \ref{fig:preprocess}.

\subsection{Constructing the Ising Hamiltonian}
For each $i$ from 0 through $n_c-1$, let ${C}_i'$ be $C_i$ after applying the classical preprocessing procedure outlined in Section~\ref{sec:simplification}. The solutions for the simplified equations ${C}'_i=0$ then correspond to the minimization of the classical energy function
\begin{equation}
E=\sum\limits_{i=0}^{n_c}{{C}'_i}^2,
\label{factoring_energy}
\end{equation}
which has a natural quantum representation as a \textit{factoring Hamiltonian}
\begin{equation}
H=\sum\limits_{i=0}^{n_c}\hat{C}_i^2.
\label{factoring_hamiltonian}
\end{equation}
Each $\hat{C}_i$ term is obtained by quantizing $p_i$, $q_i$, and $z_{j,i}$ in the clause $C_i'$ using the mapping
\begin{equation}
b_k\to\frac{1}{2}\left(1-\sigma_{b,k}^z\right),
\end{equation}
where $b\in\{p,q,z\}$ and $k$ is its associated bit index. 
We have thus encoded an instance of factoring into the ground state of a $4$-local Ising Hamiltonian. $H$ can also be represented in quadratic form by substituting each product $q_j p_{i-j}$ with a new binary variable $w_{i,j}$ and introducing additional constraints to the Hamiltonian~\cite{Dridi2017}. This is necessary for implementation on quantum annealing devices with restricted pairwise coupling between qubits. However, in our case it is not necessary since in the gate model of quantum computation methods for time evolution under $k$-local Hamiltonian are well known~\cite{Whitfield2011}.


\section{Variational quantum factoring algorithm}\label{sec:hybrid}

The main component of our scheme is an approximate quantum ground state solver for the Hamiltonian in Equation~\eqref{factoring_hamiltonian} as a means to approximately factor numbers on near-term gate model quantum computers. We use the \textit{quantum approximate optimization algorithm} (QAOA), which is a hybrid classical/quantum algorithm for near-term quantum computers that approximately solves classical optimization problems~\cite{Farhi2014}. The goal of the algorithm is to satisfy (i.e. find the simultaneous zeros of) the simplified clauses $C_i'$, which we cast as the minimization of a classical cost Hamiltonian $H_c$, and set to be identical to the Hamiltonian in Equation~\eqref{factoring_hamiltonian} (i.e. $H_c=H$).

To prepare the (approximate) ground state we use an ansatz state
\begin{equation}
\ket{\bm{\beta},\bm{\gamma}}=\prod_{i=1}^s\left(\exp\left(-i\beta_i H_a\right)\exp\left(-i\gamma_i H_c\right)\right)\ket{\bm{+}}^{\otimes n},
\label{qaoa}
\end{equation}
parametrized by angles $\bm{\beta}$ and $\bm{\gamma}$ over $n$ qubits, where $s$ is the number of layers of the QAOA algorithm. Here, $H_a$ is the \textit{admixing Hamiltonian}
\begin{equation}
H_a=\sum\limits_{i=1}^n \sigma_i^x.
\end{equation}
For a fixed $s$, QAOA uses a classical optimizer to minimize the cost function
\begin{equation}
M\left(\bm{\beta},\bm{\gamma}\right)=\bra{\bm{\beta},\bm{\gamma}}H_c\ket{\bm{\beta},\bm{\gamma}}.
\label{cost}
\end{equation}
For $s\to\infty$, $M\left(\bm{\beta},\bm{\gamma}\right)$ is minimized when the fidelity between $\ket{\bm{\beta},\bm{\gamma}}$ and the true ground state tends to $1$. Generically for $s<\infty$, $\ket{\operatorname{arg\,min}\left(M\left(\bm{\beta},\bm{\gamma}\right)\right)}$ may have exponentially small overlap with the true ground state. In our case, numerical evidence which will be discussed in Section~\ref{sec:sim} suggests that often letting $s\in O\left(n\right)$ suffices for large overlap with the ground state.



To optimize the QAOA parameters $\bm{\beta}$ and $\bm{\gamma}$, we employed a layer-by-layer iterative brute-force grid search over each pair $\left(\gamma_i,\beta_i\right)$, with the output fed into a BFGS global optimization algorithm~\cite{BFGS}. The choices for grid sizes were motivated by a gradient bound given in~\cite{Farhi2014}; more precisely, we expect each dimension of the grid should be $O\left(n_c^2 n^4\right)$. 
From~\cite{Farhi2014} a bound of $O\left(m^2+mn\right)$ is given for QAOA minimizing an objective function of $m$ clauses on $n$ variables. The setting in~\cite{Farhi2014} is that each clause gives rise to a term in the Hamiltonian that has a norm at most $1$. In our case, each clause $C_i'$ instead gives rise to a term in the Hamiltonian that has norm $\left\lvert\hat{C}_i^2\right\rvert=O\left(n^2\right)$. Therefore, we take $m=n_c n^2$ and $n$ be the number of qubits, yielding a bound $O\left(n_c^2n^4\right)$ for the gradient. To ensure that the optimum found by grid search differs from the true optimum by a constant, we therefore introduce a grid of size $O\left(n_c^2 n^4\right)\times O\left(n_c^2 n^4\right)$ based on the gradient bound~\footnote{Consider a 1D example: Suppose we would like to minimize a function $f\left(x\right)$ for $x$ on a finite interval. The derivative of the function is bounded $\left\lvert f'\left(x\right)\right\rvert<D$. Let $x^*$ be the optimal point on the interval. We introduce a mesh to discretize the interval such that the mesh point $\tilde{x}$ closest to $x^*$ has Taylor expansion $f\left(x^*\right)\approx f\left(\tilde{x}\right)+\left(\tilde{x}-x^*\right)f'\left(\tilde{x}\right)$. Then $\left\lvert f\left(x^*\right)-f\left(\tilde{x}\right)\right\rvert\le C$ for some constant $C$ when we set the mesh dense enough such that $\left\lvert x^*-\tilde{x}\right\rvert\le\frac{C}{D}$. This translates to $O\left(D\right)$ mesh points. For a 2D plane naturally the mesh choice is $O\left(D\right)\times O\left(D\right)$.}. 
This ensures a polynomial scaling of the grid resolution. Numerically, training on coarser grids seemed sufficient (see Table~\ref{table:numbers}).

\begin{table}
\begin{tabular}{|c|c|c|c|c|}
\hline
Input number $m$ & Number of qubits $n$ & Number of carry bits & $p\leftrightarrow q$ symmetry & Grid size \\
\hline
$35 = 5\times 7$& $2$ & $0$ & \ding{51} & $6\times 6$ \\
$77 = 7\times 11$ & $6$ & $3$ & \ding{55} & $24\times 24$\\
$1207 = 17\times 71$ & $8$ & $5$ & \ding{55} & $36\times 36$\\
$33667 = 131\times 257$ & $3$ & $1$ & \ding{55} & $9\times 9$\\
$56153 = 233\times 241$ & $4$ & $0$ & \ding{51} & $12\times 12$\\
$291311 = 523\times 557$  & $6$ & $0$ & \ding{51} & $24\times 24$\\
\hline
\end{tabular}
\caption{First column: Biprime numbers used in this study. Second column: the total number of qubits needed to perform VQF on the problem instance. Third column: among the qubits, the number of carry bits produced in the Ising Hamiltonian after simplifying the Boolean equations with rules described in~\eqref{boolean_subs}. The observed difference between instances with carry bits versus without carry bits is shown in Figure~\ref{fig:layers}, along with Figures~\ref{fig:factoring_with_carries} and~\ref{fig:factoring}. Fourth column: in the energy function~\eqref{factoring_energy}, whether or not there exists a $p\leftrightarrow q$ symmetry. Such symmetry can be broken by two factors having different bit lengths. Fifth column: size of the grid used for the layer-by-layer brute-force search.}
\label{table:numbers}
\end{table}

\begin{figure}[ht]
\resizebox{\linewidth}{!}{
\includegraphics{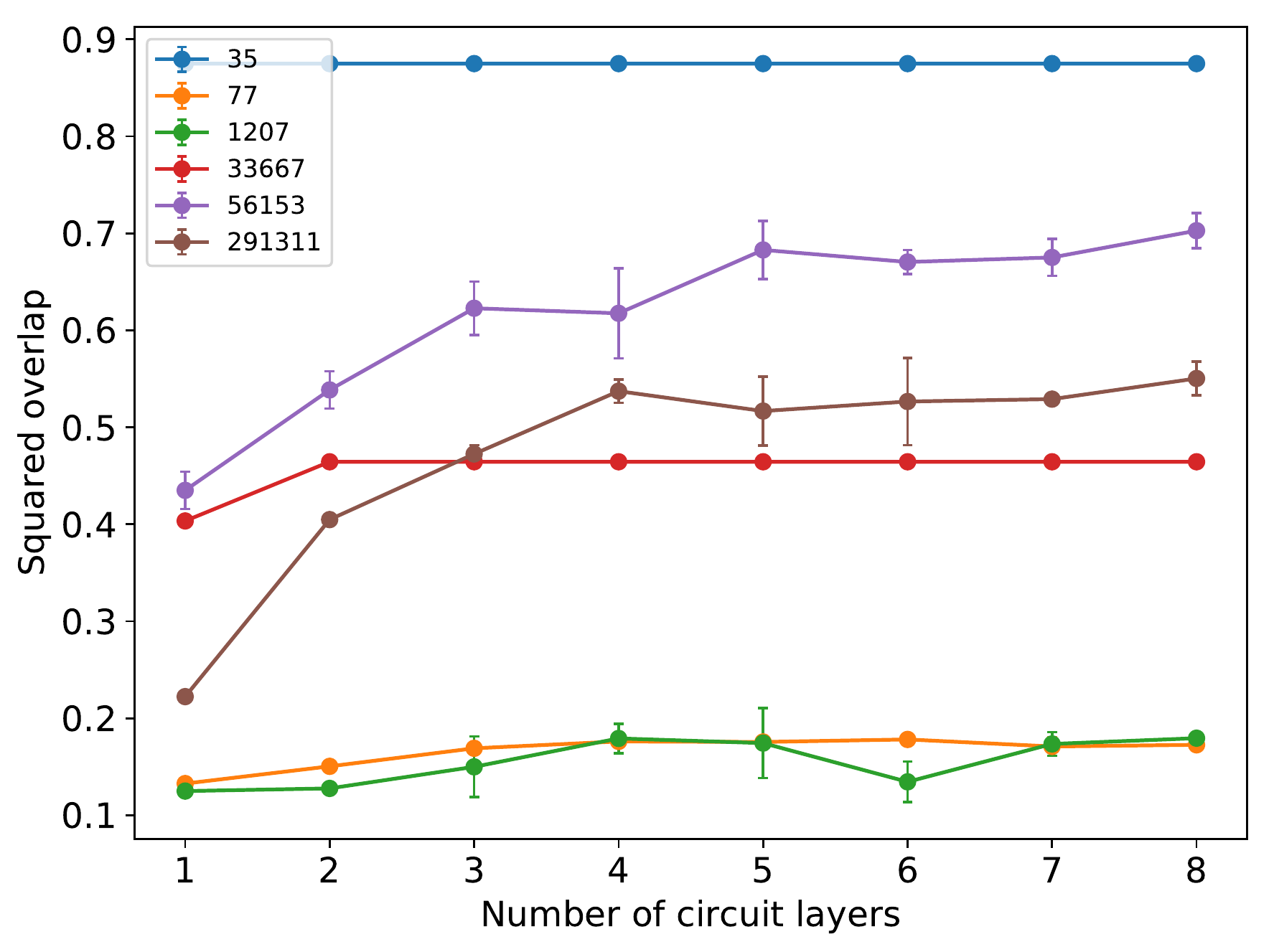}
}
\caption{The squared overlap of the optimized VQF state with the solution state manifold of $H_c$ for all problem instances considered. Here, we fixed the error rate $\varepsilon=10^{-3}$ and the number of samples $\nu=10000$. We note the drastically reduced depth scaling for $m=77,1207,33667$ (see Section~\ref{sec:scaling}). The error bars each denote one standard deviation over three problem instances.
}
\label{fig:layers}
\end{figure}

\begin{figure}[ht]
{\resizebox{\linewidth}{!}{\includegraphics{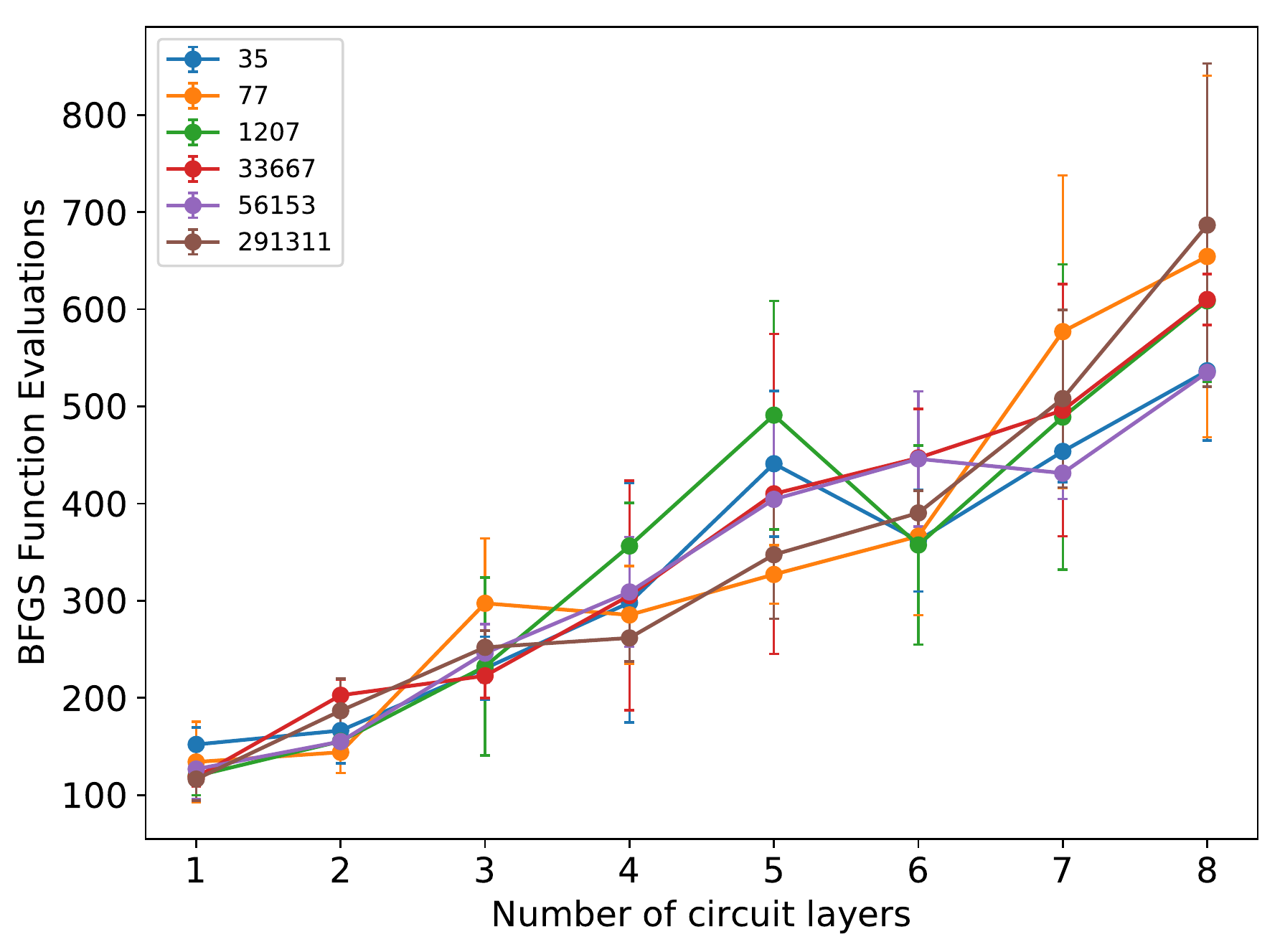}}}
\caption{The scaling of the number evaluations of Equation~\eqref{cost} needed before the BFGS optimization converges. The scaling is approximately linear in the number of parameters, and is approximately independent of the problem size. The error bars each denote one standard deviation over three problem instances.
}
\label{fig:bfgs_evals}
\end{figure}

The remaining cost for finding the solution then comes from the global optimization procedure. In our numerical studies, the complexity scaling of performing BFGS optimization until convergence (to either a local or a global minimum) seemed independent of the problem size and depended linearly on the circuit depth (see Figure~\ref{fig:bfgs_evals}).  For a QAOA ansatz of depth $s$, this puts the total cost of performing VQF at $O\left(s^2 n_c^4 n^8\right)$ in the worst case, though numerically, this seems like a loose bound. We also note that there is no guarantee that this procedure always generates the globally optimal solution.

\section{Numerical Simulations}\label{sec:sim}
\subsection{Depth Scaling}\label{sec:scaling}
We performed noisy simulation of a number of instances of biprime factoring using the algorithm described above~\footnote{The simulation is performed using QuTiP~\cite{Qutip}. To access the data generated for all instances considered in this study, including those which produced Figures~\ref{fig:layers}-\ref{fig:factoring_with_carries}, please refer to our Github repository at \url{https://github.com/zapatacomputing/VQFData}.} (see Section~\ref{sec:noise} for a description of our noise model).
Table~\ref{table:numbers} lists all of the instances used. With the technique described in Section~\ref{sec:hybrid}, the success probability of finding the correct factors of $m=35,77,1207,33667,56153,291311$ as a function of the number of circuit layers $s$ is plotted in Figure~\ref{fig:layers}. The output distributions for representative numbers are plotted in Figures~\ref{fig:factoring} and~\ref{fig:factoring_with_carries}. Here, ``squared overlap'' refers to the squared overlap of the output VQF state with the solution state manifold of $H_c$---that is, the squared overlap with states with the correct assignments of all $p_i$ and $q_i$ but not necessarily of all the carry bits $z_{ij}$, which are not bits of the desired factors $p$ and $q$.

For $m=35,56153,291311$, after $O\left(n\right)$ circuit layers, the success probability plateaus to a large fraction. As factoring is efficient to check, one can then sample from the optimized VQF ansatz and check samples until correct factors of $m$ are found. However, the algorithm does not scale as well with the circuit depth for $m=77,1207,33667$. This is the case even though the $m=77,33667$ problem instances have the same number or fewer qubits required than the $m=56153,291311$ problem instances. Further insight is needed to explain this discrepancy, though we do notice that unlike $m=35,56153,291311$, these instances lack $p\leftrightarrow q$ symmetry and contain carry bits in their classical energy functions~\eqref{factoring_energy} (see Table~\ref{table:numbers}).


\subsection{Noise Scaling}\label{sec:noise}
\begin{figure*}[ht]
\begin{subfloat}[$m=56153$]{\resizebox{0.5\linewidth}{!}{\includegraphics{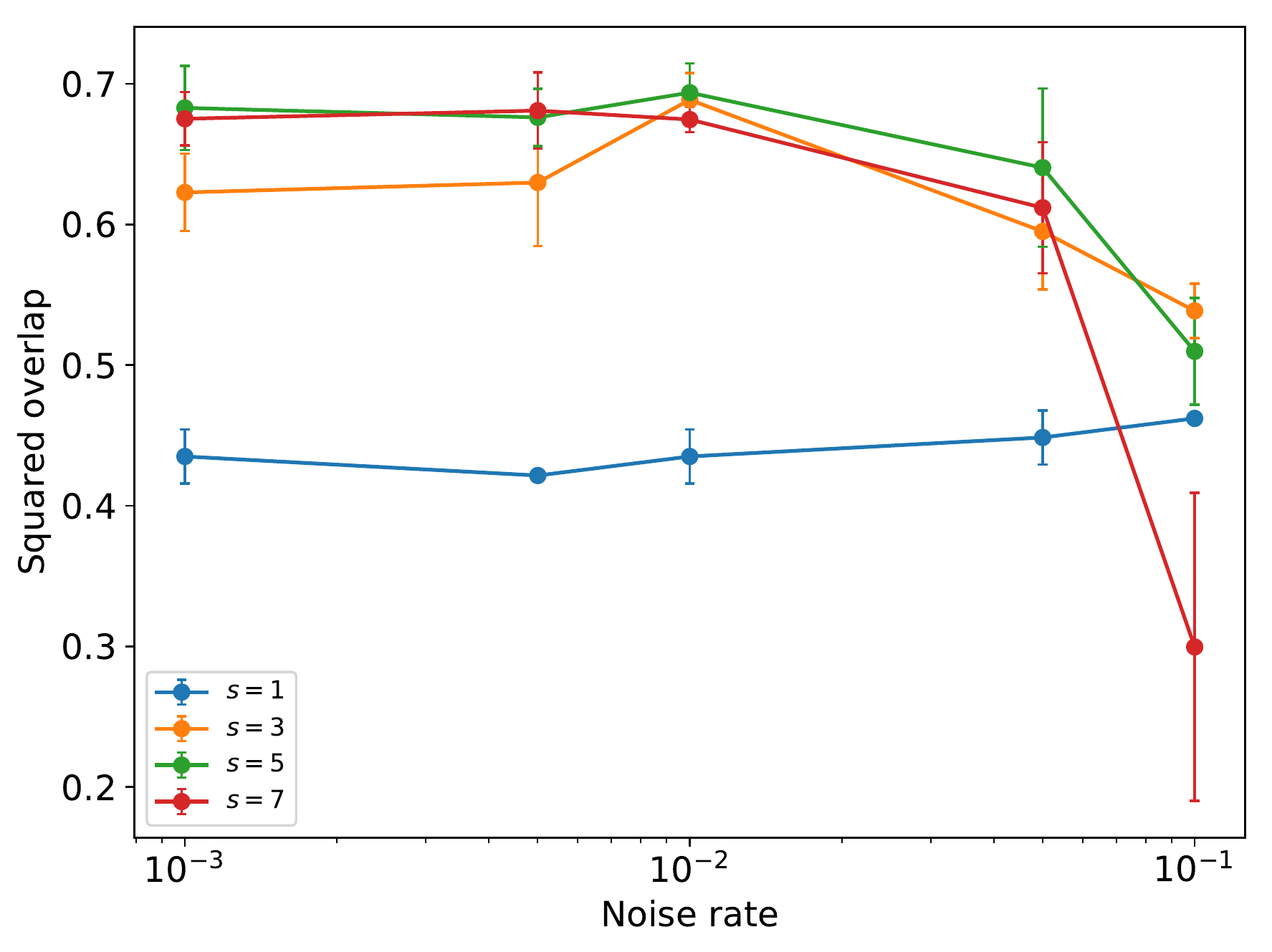}}}\end{subfloat}\begin{subfloat}[$m=77$]{\resizebox{0.5\linewidth}{!}{\includegraphics{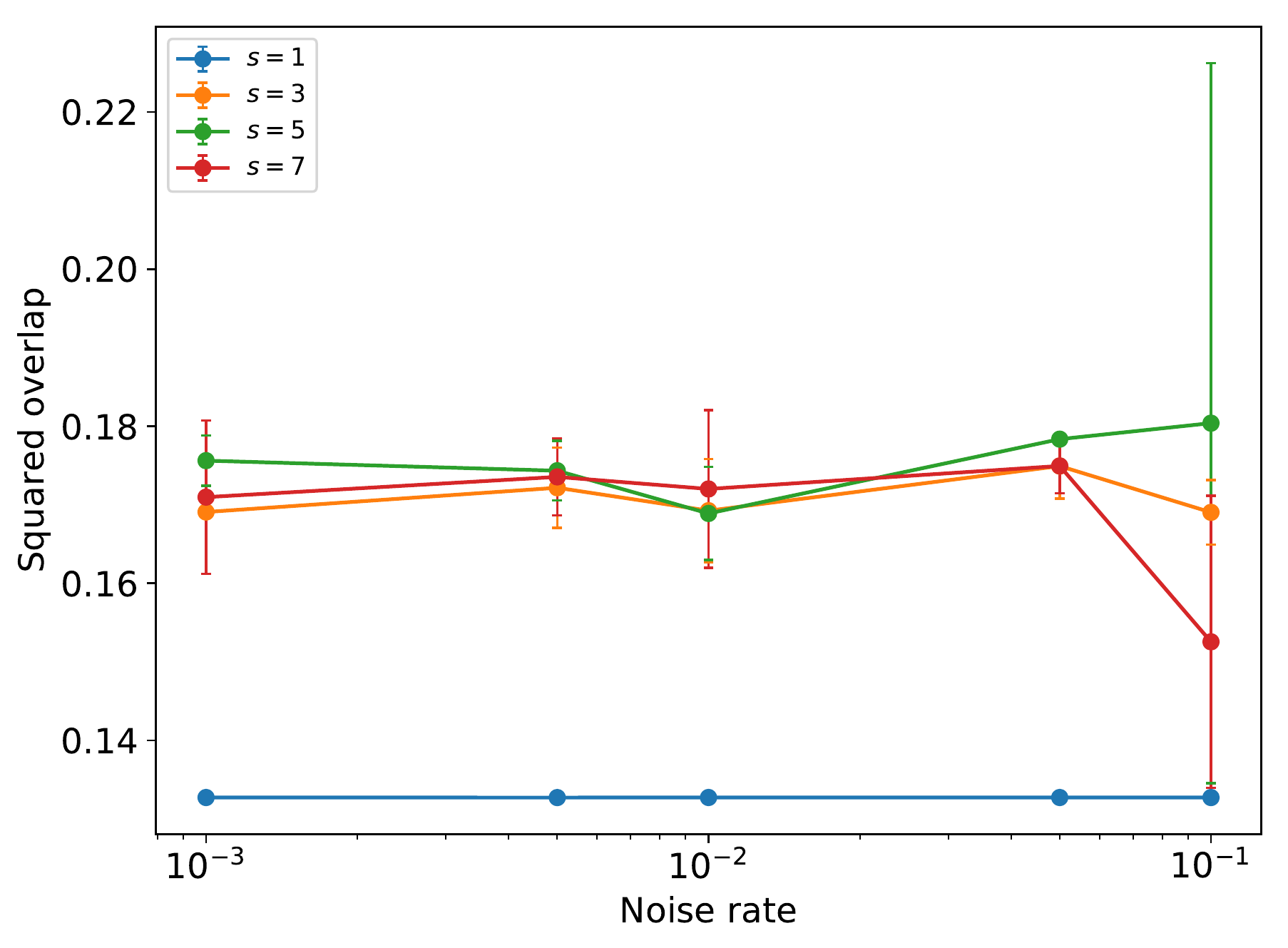}}}\end{subfloat}
\caption{The dependence on factoring (a) $m=56153$ and (b) $m=77$ at various depths for different Pauli error noise rates. Below a certain error threshold, the success probability is approximately independent of the noise rate. The error bars each denote one standard deviation over three problem instances.}
\label{fig:noise}
\end{figure*}

\begin{figure*}[ht]
\begin{subfloat}[$s=1$, $m=56153$]{\resizebox{0.425\linewidth}{!}{\includegraphics{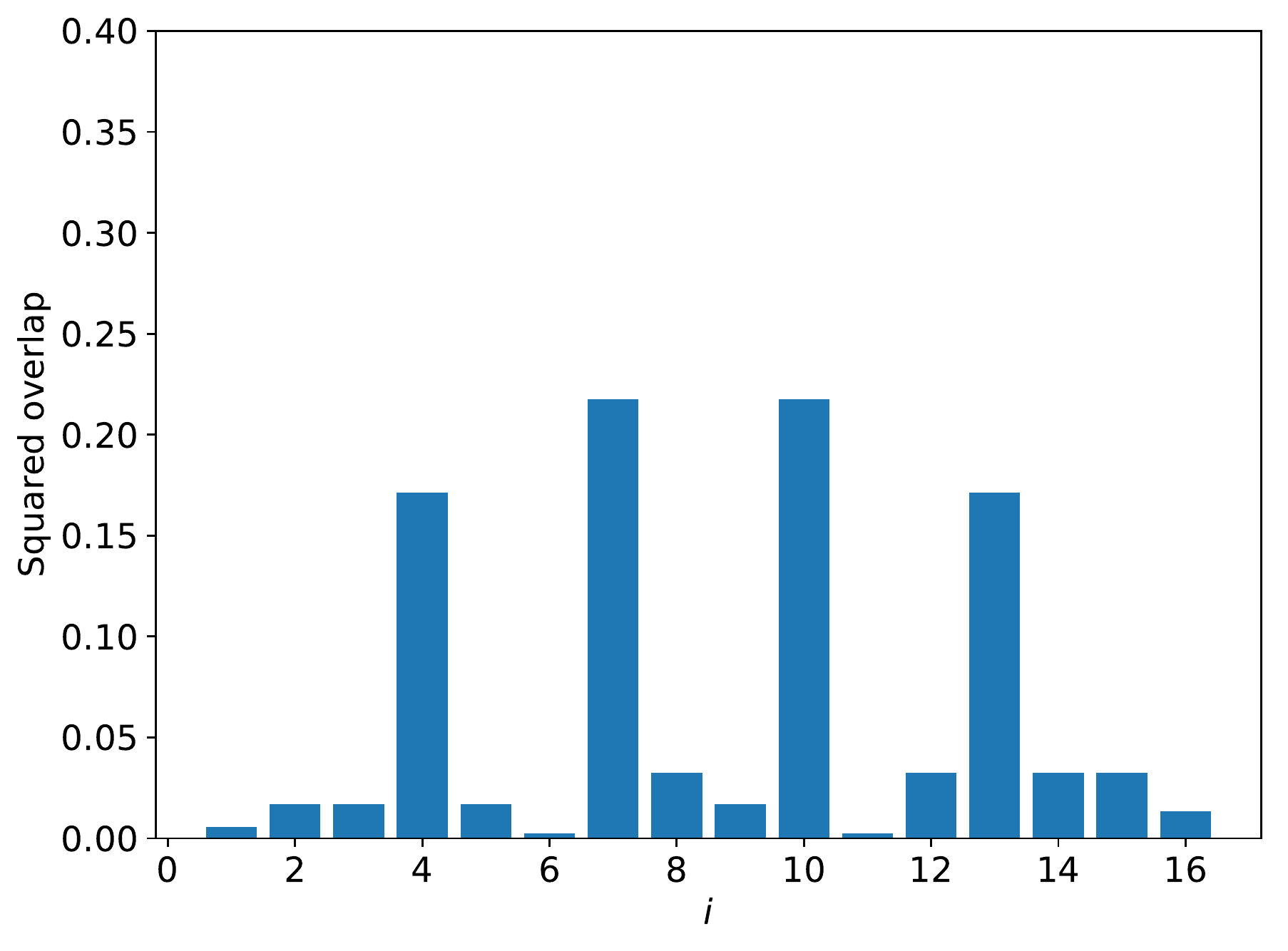}}}\end{subfloat}\begin{subfloat}[$s=2$, $m=291311$]{\resizebox{0.425\linewidth}{!}{\includegraphics{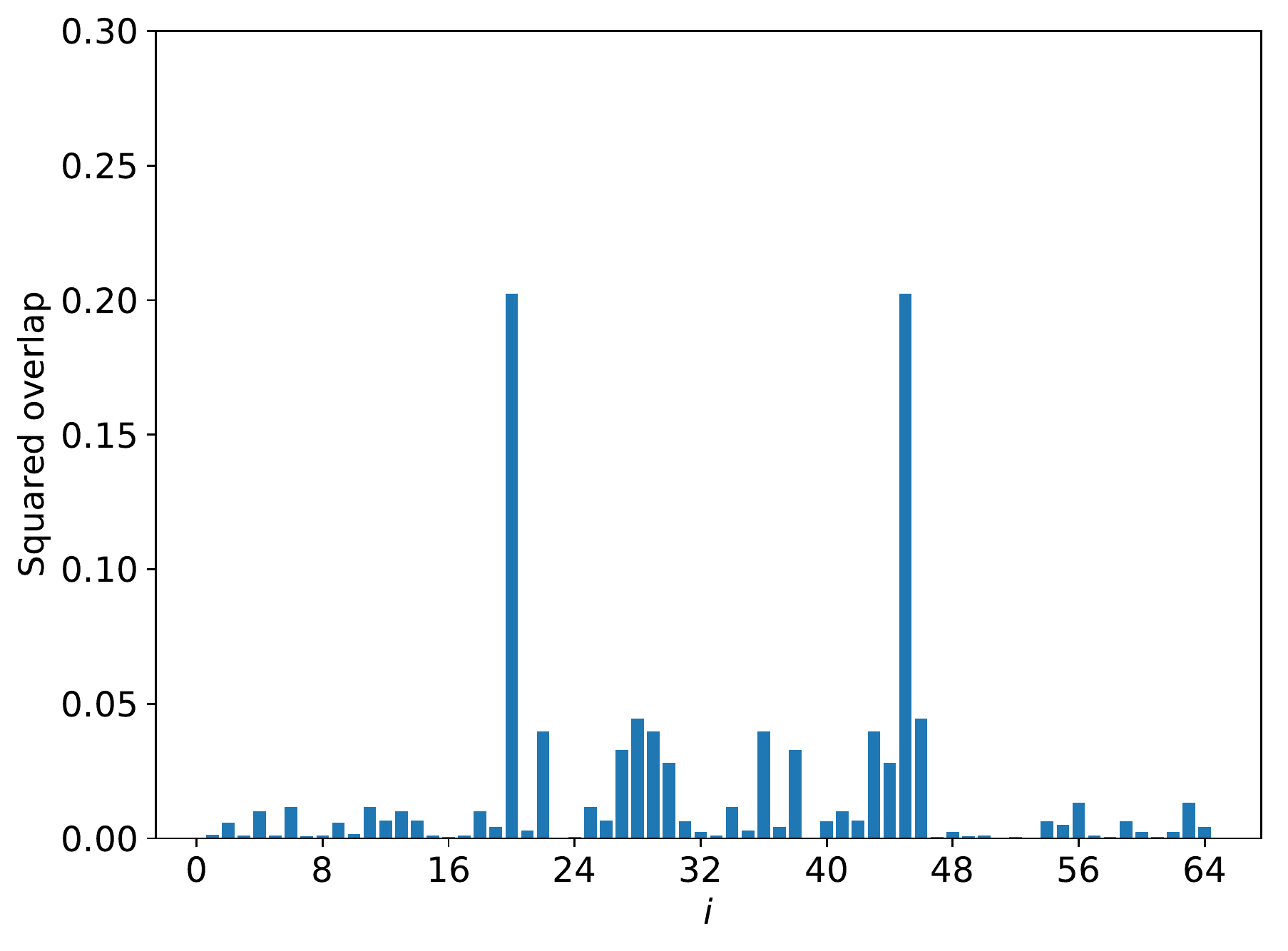}}}\end{subfloat}
\begin{subfloat}[$s=2$, $m=56153$]{\resizebox{0.425\linewidth}{!}{\includegraphics{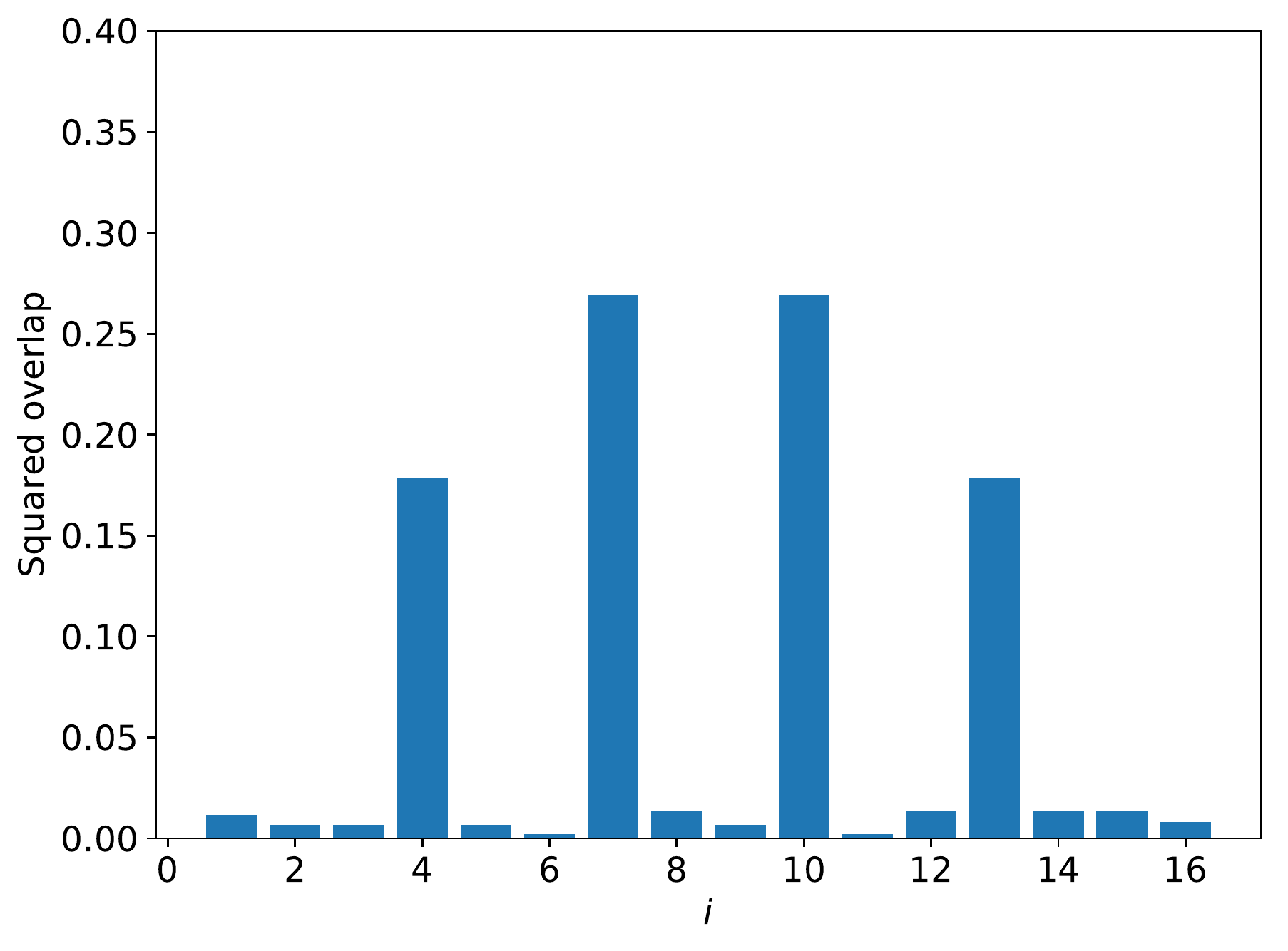}}}\end{subfloat}\begin{subfloat}[$s=4$, $m=291311$]{\resizebox{0.425\linewidth}{!}{\includegraphics{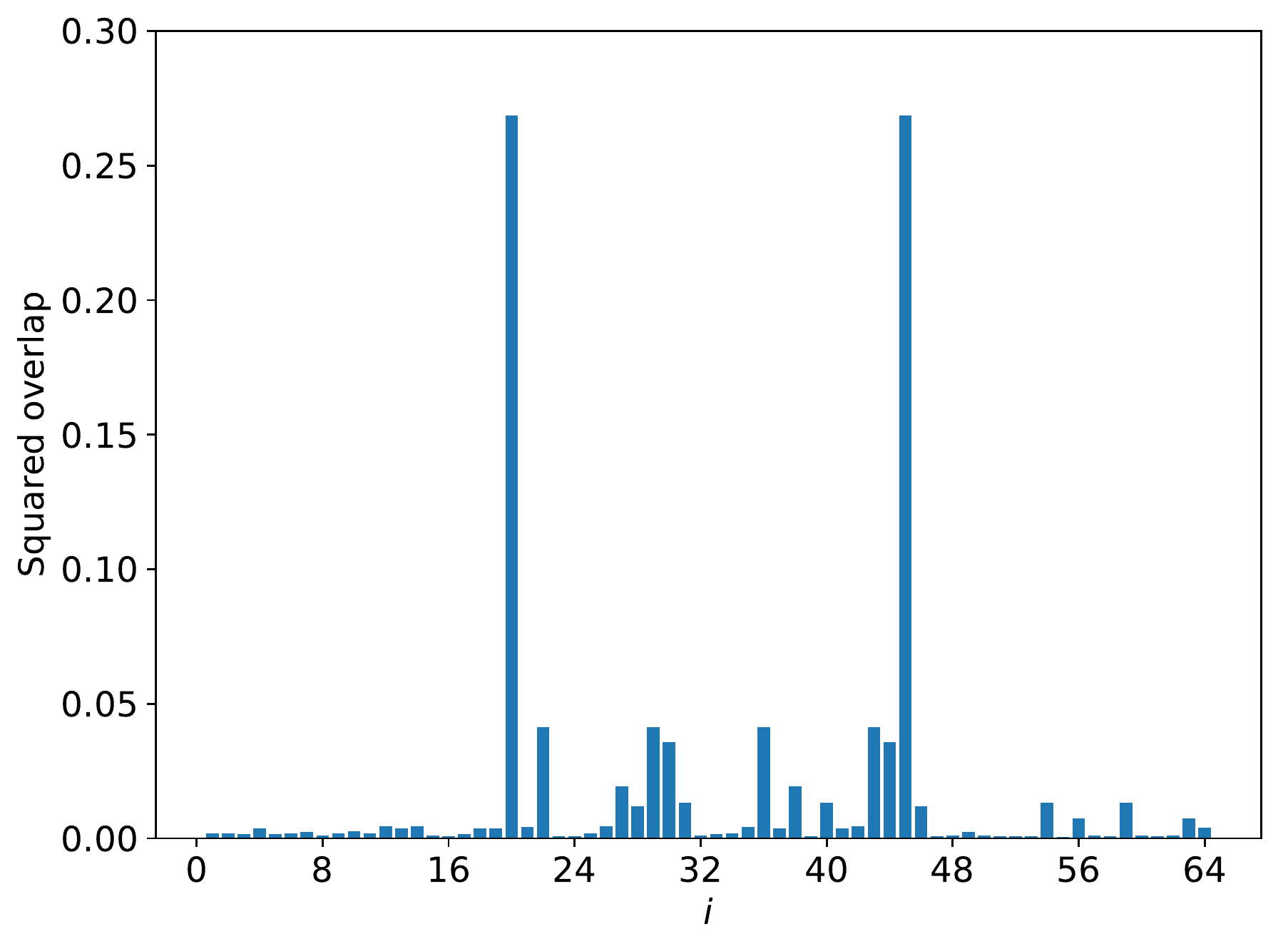}}}\end{subfloat}
\begin{subfloat}[$s=3$, $m=56153$]{\resizebox{0.425\linewidth}{!}{\includegraphics{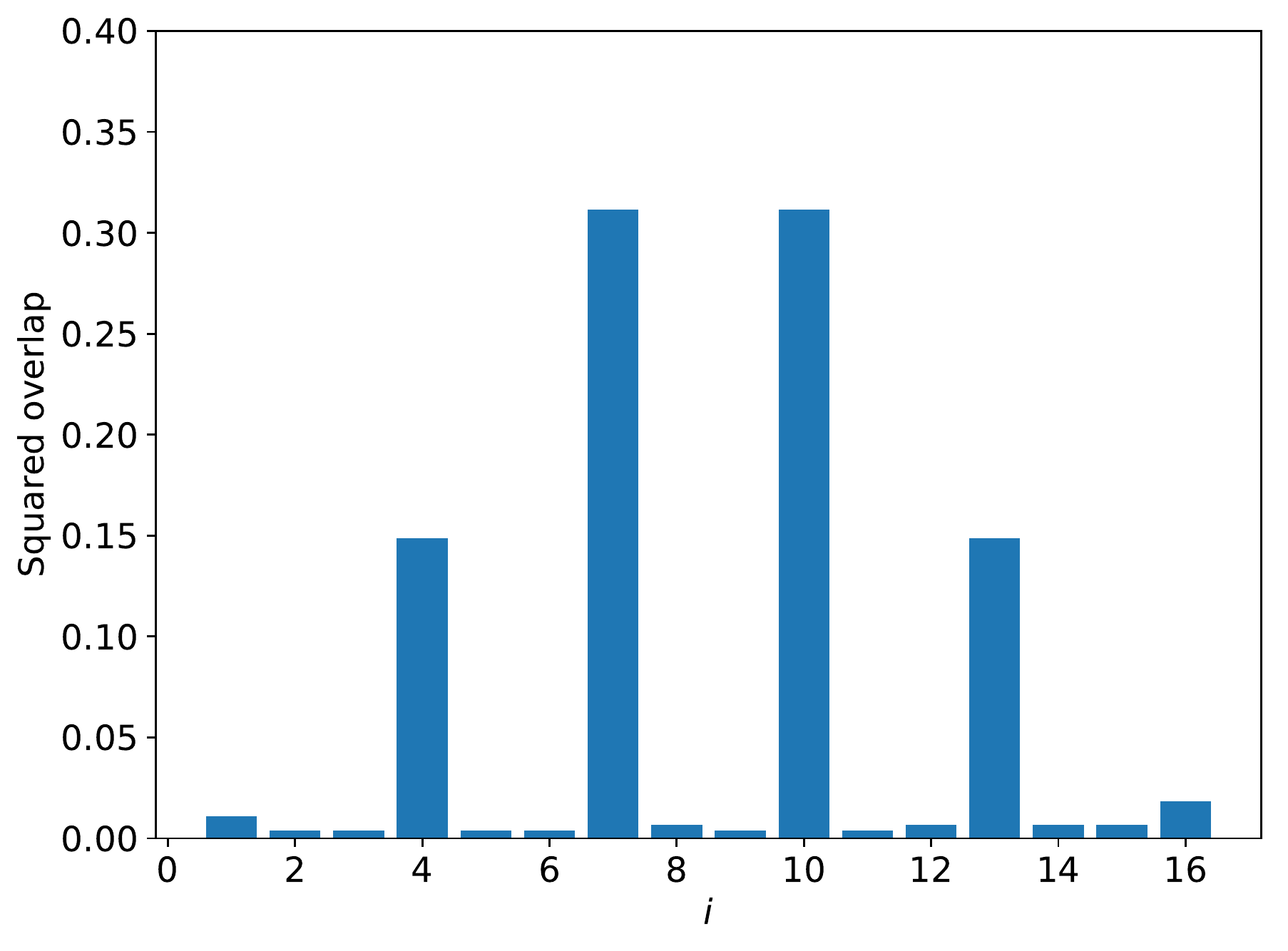}}}\end{subfloat}\begin{subfloat}[$s=6$, $m=291311$]{\resizebox{0.425\linewidth}{!}{\includegraphics{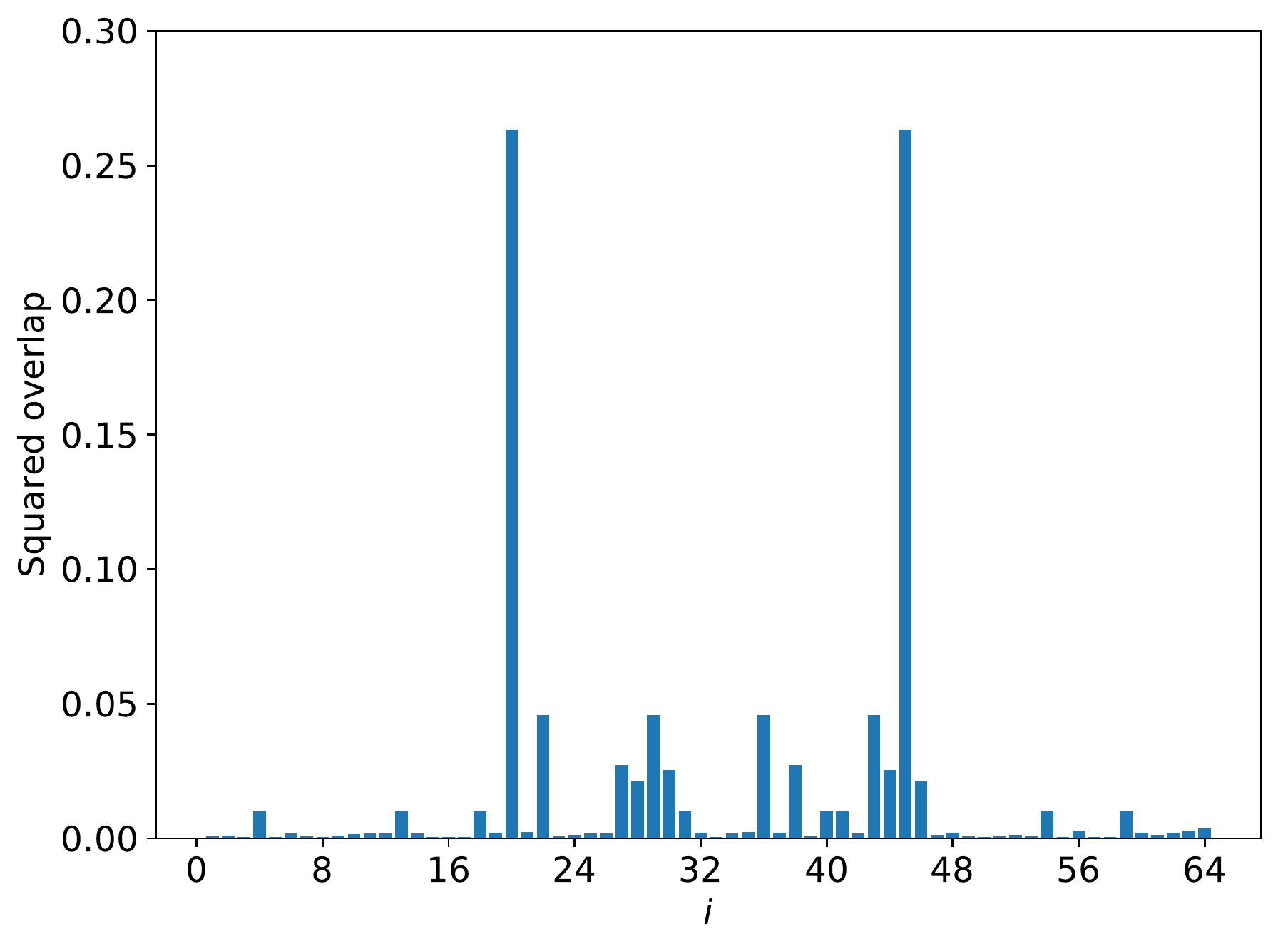}}}\end{subfloat}
\caption{Distributions corresponding to the output of the presented factoring algorithm for various circuit depths. $i$ labels computational basis states in lexicographic order. The two modes of each diagram correspond to the computational basis states yielding the correct $p$ and $q$; there are two modes due to the $p\leftrightarrow q$ symmetry of the problem. Here, we fixed the error rate $\varepsilon=10^{-3}$ and the number of samples $\nu=10000$. 
}
\label{fig:factoring}
\end{figure*}

\begin{figure*}[ht]
\begin{subfloat}[$s=1$, $m=77$]{\resizebox{0.425\linewidth}{!}{\includegraphics{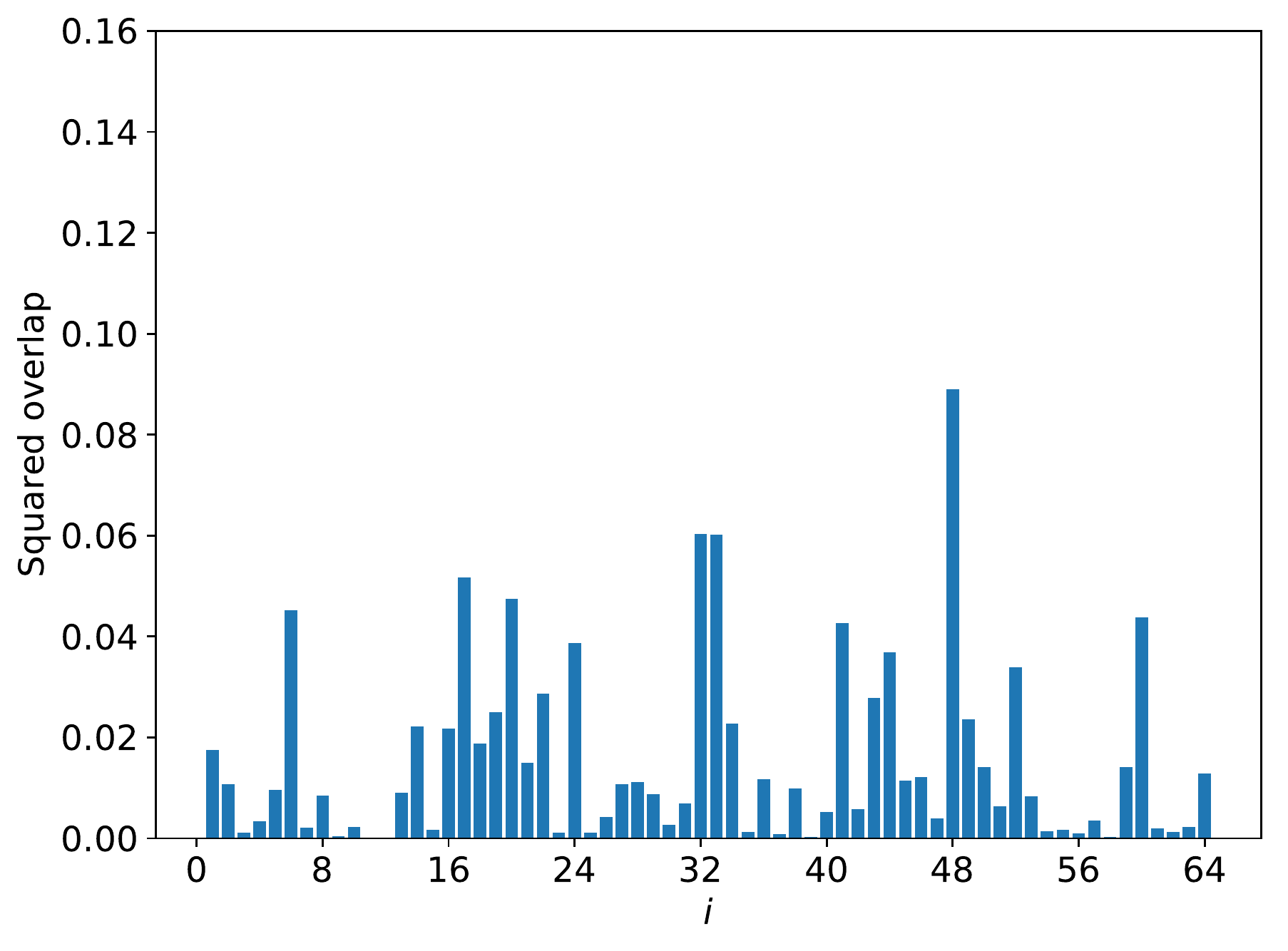}}}\end{subfloat}\begin{subfloat}[$s=1$, $m=1207$]{\resizebox{0.425\linewidth}{!}{\includegraphics{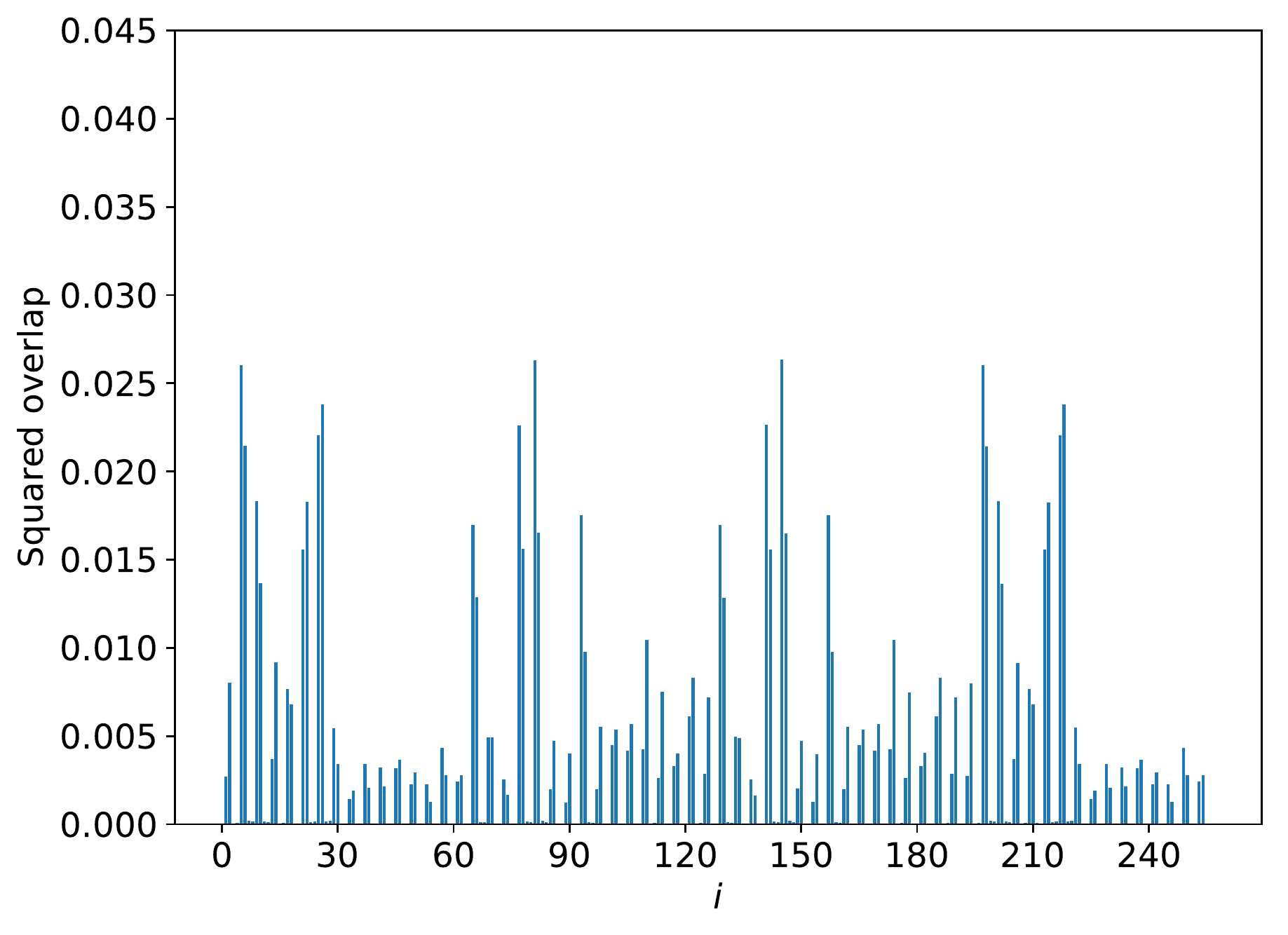}}}\end{subfloat}
\begin{subfloat}[$s=4$, $m=77$]{\resizebox{0.425\linewidth}{!}{\includegraphics{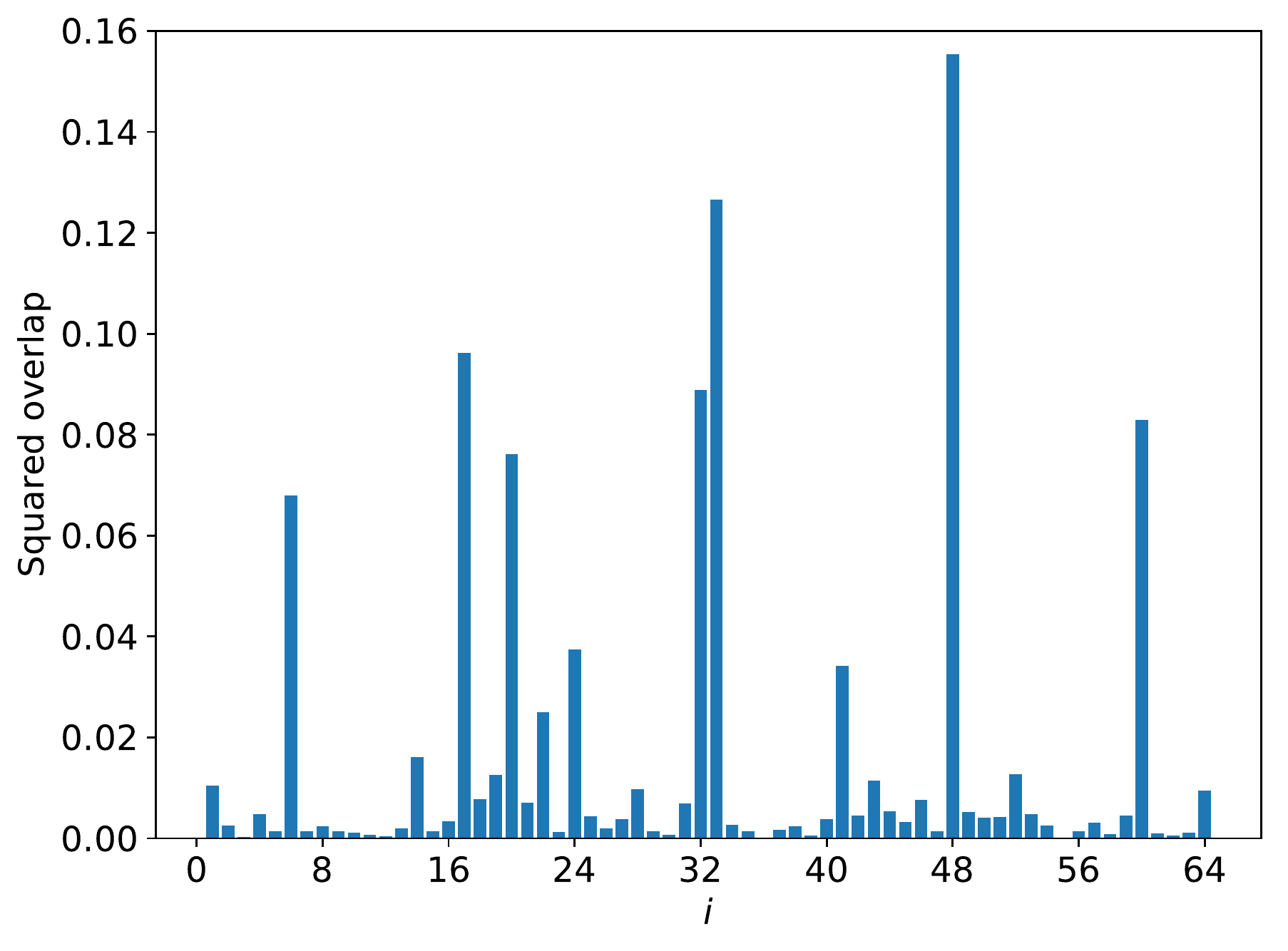}}}\end{subfloat}\begin{subfloat}[$s=4$, $m=1207$]{\resizebox{0.425\linewidth}{!}{\includegraphics{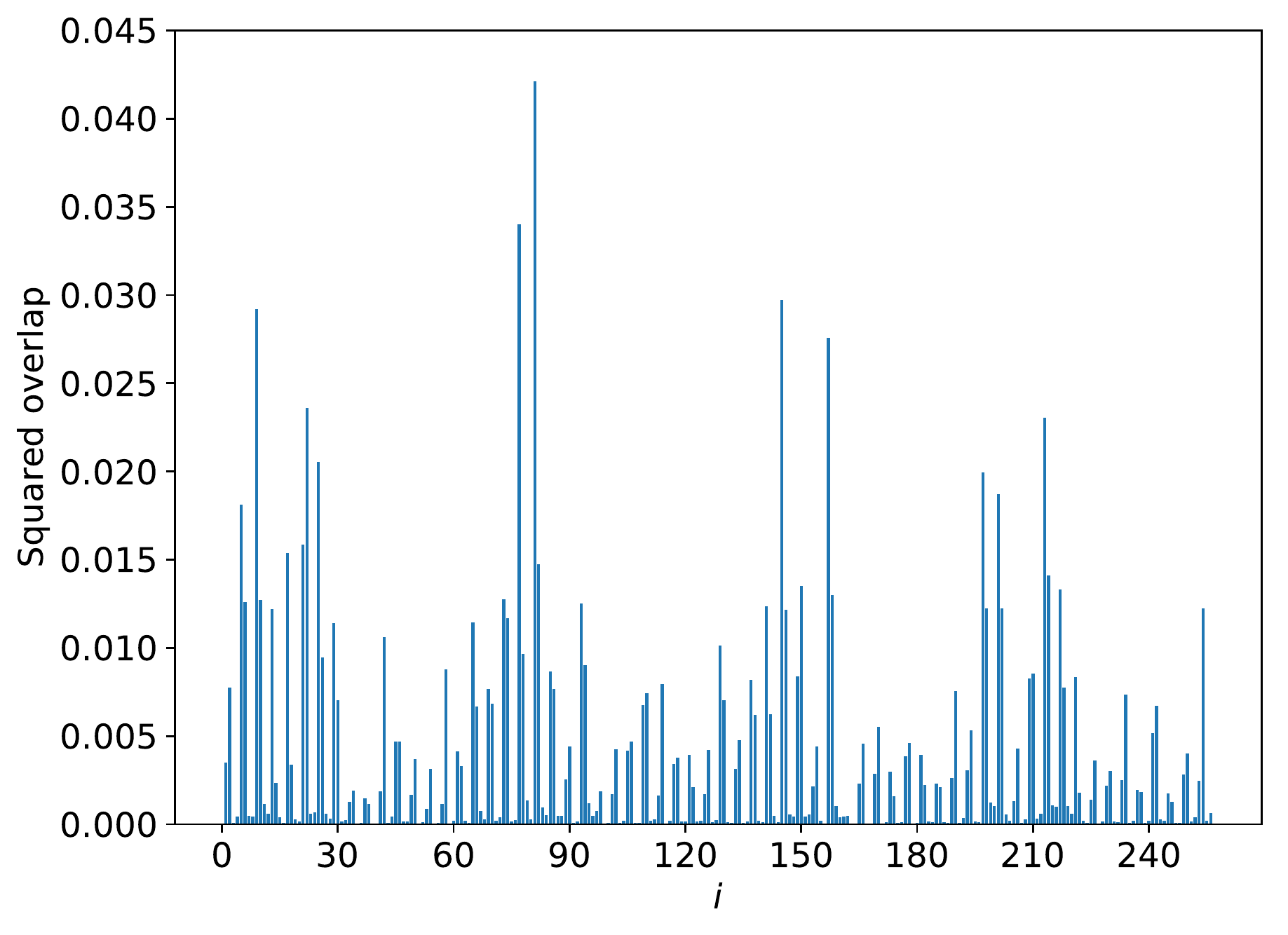}}}\end{subfloat}
\begin{subfloat}[$s=8$, $m=77$]{\resizebox{0.425\linewidth}{!}{\includegraphics{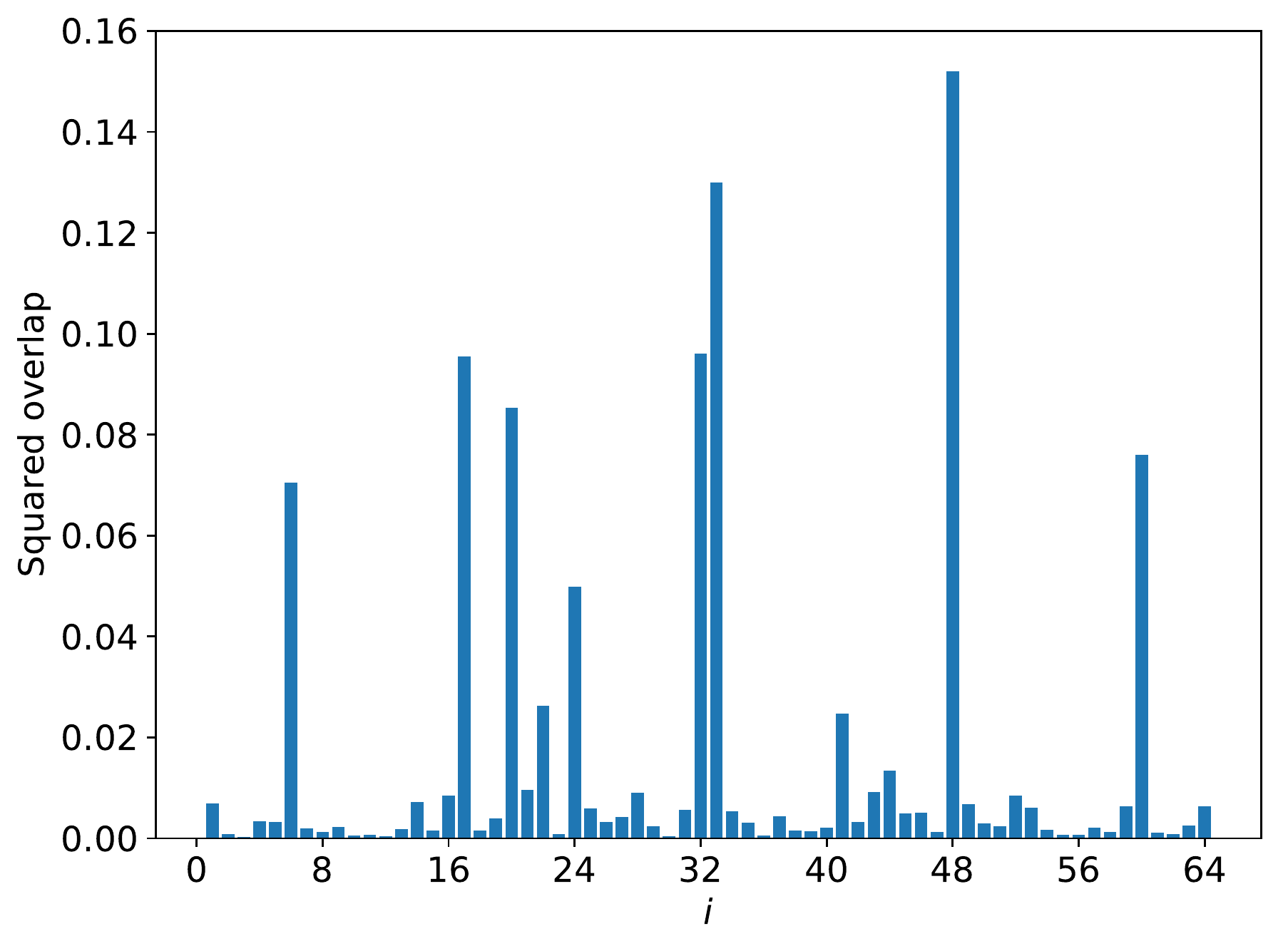}}}\end{subfloat}\begin{subfloat}[$s=8$, $m=1207$]{\resizebox{0.425\linewidth}{!}{\includegraphics{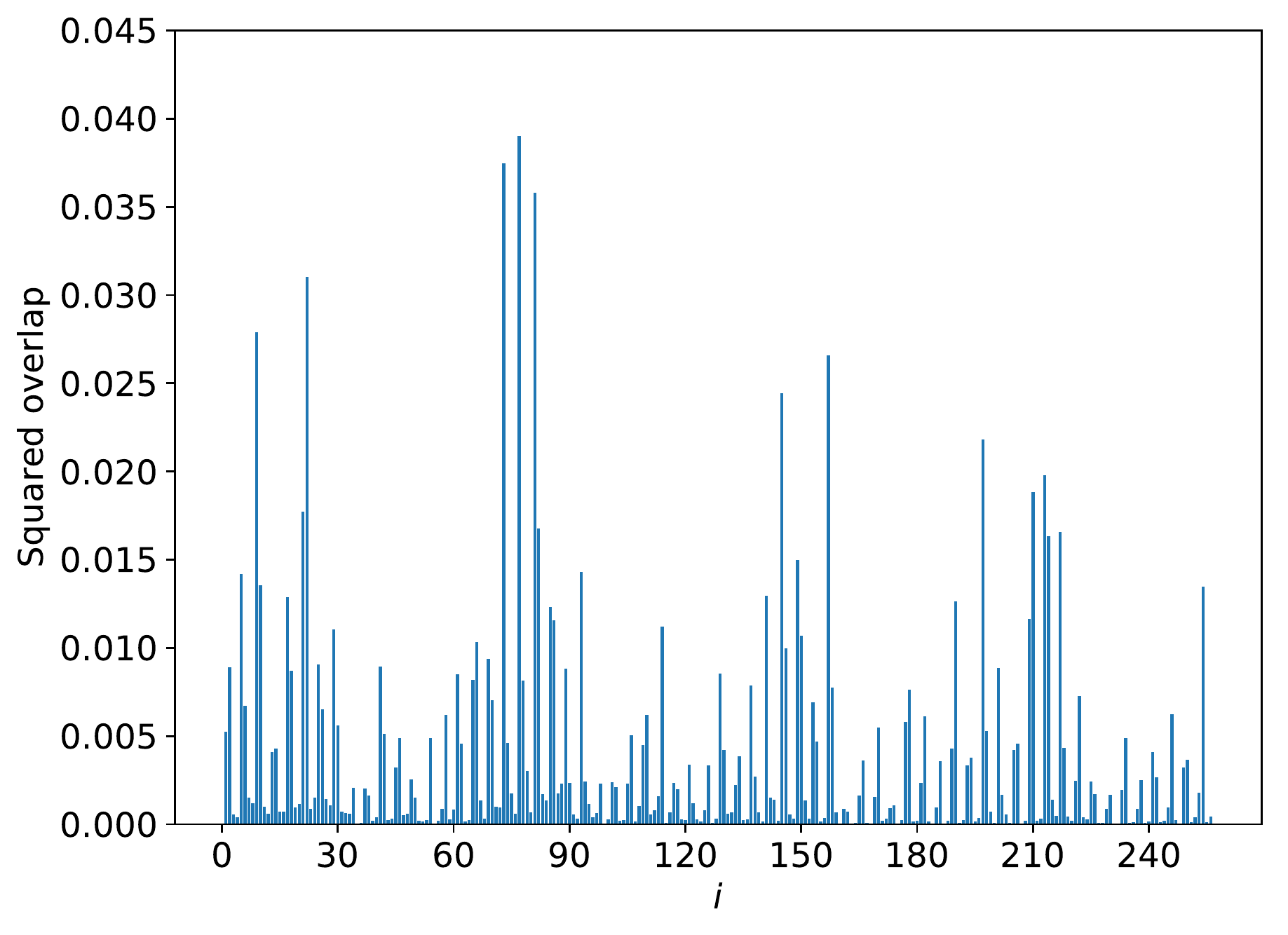}}}\end{subfloat}
\caption{Distributions corresponding to the output of the presented factoring algorithm for various circuit depths. $i$ labels computational basis states in lexicographic order. The modes of the high depth distributions are the correct ground states. We notice worse performance than $m=56153,291311$ (see Figure~\ref{fig:factoring} and Section~\ref{sec:scaling}). Here, we fixed the error rate $\varepsilon=10^{-3}$ and the number of samples $\nu=10000$.}
\label{fig:factoring_with_carries}
\end{figure*}

An obvious concern for the scalability of the algorithm is the effect of noise on the performance of VQF.  To explore this empirically, we considered a Pauli channel error model; that is, after every unitary (and after the preparation of $\ket{+}^{\otimes n}$) in Equation~\eqref{qaoa}, we implemented the noise channel
\begin{equation}
\rho\mapsto\left(1-n\varepsilon\right)\rho+\frac{\varepsilon}{3}\sum\limits_{j=1}^n\sum\limits_{i=1}^3\sigma_j^{\left(i\right)}\rho\sigma_j^{\left(i\right)},
\end{equation}
where $\varepsilon$ is the single qubit error rate. Included in the simulation is sampling noise with $\nu=10000$ samples when estimating the cost function $M\left(\bm{\beta},\bm{\gamma}\right)$. We plot the dependence of two VQF instances on the noise rate in Figure~\ref{fig:noise}, and note that VQF is weakly dependent on the noise rate below a certain error threshold.

\section{Discussion}\label{sec:discuss}



The ability to efficiently solve integer factorization has significant implications for public-key cryptography. In particular, encryption schemes based on abelian groups such as RSA and elliptic curves can be compromised if efficient factorization were feasible. However, an implementation of Shor's algorithm for factoring cryptographically relevant integers would require thousands of \textit{error-corrected} qubits~\cite{RodneyVanMeterThaddeusD.LaddAustinG.Fowler2010DistributedNanophotonics,Jones2012LayeredComputing}. This is far too many for noisy intermediate-scale quantum devices that are available in the near-term, rendering the potential of quantum computers to compromise modern cryptosystems with Shor's algorithm a distant reality. Hybrid approximate classical/quantum methods that utilize classical pre- and post-processing techniques, like the proposed VQF approach, may be more amenable to factoring on a quantum computer in the next decade.

Although we show that it is in principle possible to factor using VQF, as with most heuristic algorithms, it remains to be seen whether it is capable of scaling asymptotically under realistic constraints posed by imperfect optimization methods and noise on quantum devices.  We are currently in the process of examining more detailed analytical and empirical arguments to better determine the potential scalability of the protocol under realistic NISQ conditions. We look forward to working with our collaborators on experimental implementations on current NISQ devices.

The VQF approach can also be employed in an error-corrected setting. Given its heuristic approach it presents a tradeoff between the number of coherent gates and the number of repetitions, similar to the previous VQE and QAE approaches. In this sense, VQF could be competitive with Shor's algorithm even in the regime of fault-tolerant quantum computation. However, further work is needed in comparing the resources needed for both approaches, including understanding what causes VQF to struggle with certain factoring instances---preliminary numerics suggest that the mere presence of carry bits negatively affects the algorithm, with little dependence on the number of carry bits for a fixed problem size.

In conclusion, the VQF approach discussed here presents many stimulating challenges for the community. QAOA, the optimization algorithm employed in our approach, has been studied by several groups in order to understand its effectiveness in several situations~\cite{Farhi2014,Farhi2016QuantumAlgorithm,Nannicini2018PerformanceOptimization,Venturelli2017TemporalCircuits,Lin2016PerformanceDegree,Otterbach2017UnsupervisedComputer,Otterbach2017UnsupervisedComputer,Fried2017QTorch:Handler}. VQF inherits both the power and limitations of QAOA, and therefore many more numerical and analytical studies are needed to understand the power of VQF in the near future.

\begin{acknowledgments}
We would like to acknowledge the Zapata Computing scientific team, including Peter Johnson, Jhonathan Romero, Borja Peropadre, and Hannah Sim for their insightful and inspiring comments.
\end{acknowledgments}

\bibliography{main.bib}

\end{document}